\documentclass[11pt,a4paper]{article}

\usepackage{epsfig}
\usepackage{graphicx}
\usepackage{amsfonts}
\usepackage{amssymb}
\usepackage{bm}
\usepackage{latexsym}
\usepackage{amsmath}
\usepackage{hyperref}
\usepackage{xcolor}

\newcommand{\beq}{\begin{equation}}
\newcommand{\eeq}{\end{equation}}
\newcommand{\bea}{\begin{eqnarray}}
\newcommand{\eea}{\end{eqnarray}}

\begin{document}
%%%%%%%%%%%%%%%%%%%%%%%%%%%%%%%%%%%%%%%%%%%%%%%%%%%%%%%%%%%%%%%%%%%%
\begin{titlepage}

\title{\Large \bf Leading Gravitational Corrections and A Unified Universe}

\author{Alessandro Codello and Rajeev Kumar Jain\footnote{Corresponding author}}
\date{\normalsize  CP$^3$-Origins, Centre for Cosmology and Particle Physics Phenomenology \\
University of Southern Denmark\\
Campusvej 55, 5230 Odense M, Denmark \medskip \\
E-mail: codello@cp3-origins.net, jain@cp3.sdu.dk}

\maketitle

%%%%%%%%%%%%%%%%%%%%%%%%%%%%%%%%%%%%%%%%%%%%%%%%%%%%%%%%%%%%%%%%%%%%
\bigskip
\noindent {\large\bf Abstract } \bigskip \\
Leading order gravitational corrections to the Einstein--Hilbert action can lead to a consistent picture of the universe by unifying the epochs of inflation and dark energy in a single framework. While the leading local correction induces an inflationary phase in the early universe, the leading non--local term leads to an accelerated expansion of the universe at the present epoch. We argue that both the leading UV and IR terms can be obtained within the framework of a covariant effective field theory of gravity. The perturbative gravitational corrections therefore provide a fundamental basis for understanding a possible connection between the two epochs.
 
\thispagestyle{empty}
  \vfill 
  \flushleft{Essay written for the Gravity Research Foundation 2016 Awards for Essays on Gravitation.\\ Submitted  March 31, 2016.}
\end{titlepage}

Cosmological observations strongly suggest that our universe underwent an early period of accelerated expansion called inflation and is also experiencing a phase of acceleration at the present epoch, caused by dark energy. Whether there exists a fundamental connection between them or not, an interesting question to ask is if it is possible to unify them in a single minimal framework.  
Within General Relativity (GR) and {\sl sans} a cosmological constant, it is not possible to explain either inflation or dark energy without adding extra degrees of freedom. Therefore, these two epochs are novel consequences of physics beyond classical GR, described by the Einstein--Hilbert (EH) action.
The framework of the covariant Effective Field Theory (EFT) of quantum gravity, developed in \cite{Codello:2015mba, Codello:2015pga}, predicts both local (UV) and non--local (IR) corrections which become natural candidates for driving inflation and dark energy thereby allowing us to construct a unified picture of the universe. The key advantage of using an EFT approach is that it allows to compute such corrections from first principles even in the absence of a complete theory of quantum gravity \cite{Donoghue:1994dn,Donoghue:2015hwa}.

Besides scalar fields, inflation can also be driven by a quadratic $R^2$ term, a leading one--loop quantum correction to the EH action  -- a scenario first proposed by Starobinsky \cite{Starobinsky:1980te} and further discussed in \cite{Mukhanov:1981xt}. Being the simplest, this scenario should be utmost probable from Occam's razor and also turns out to be the most preferred model in recent datasets \cite{Martin:2013tda,Ade:2015lrj}. In the covariant EFT of gravity, the  $R^2$ term naturally arises as a leading UV term to the EH action which drives inflation at early times without the need of additional matter fields. Thus, inflation in this framework is entirely a feature of the leading gravitational corrections \cite{Codello:2015pga}.
On the contrary, the present acceleration of our cosmos can be achieved by simply adding a cosmological constant $\Lambda$.  In addition,  specific curvature square non--local terms, e.g. $R \frac{1}{\square^2} R$,  can also drive the same as it effectively behaves like a cosmological constant at late times thereby providing a viable alternative to dynamical dark energy.
An immediately relevant question to ask here is: How can one explain the presence of such an IR term in a consistent framework ?
Covariant EFT once again comes to rescue. The computation of one--loop quantum corrections leads to various non--local terms at the second order in curvatures \cite{Codello:2015mba, Codello:2015pga}. However,  $R \frac{1}{\square^2} R$ term among others turns out to be the most relevant IR term responsible for dark energy.

This essay therefore presents a unified framework to describe the evolution of the universe from very early times until the present epoch by including both the leading UV $R^2$ and leading IR $R\frac{1}{\square^2}R$ corrections to the EH action together with radiation and matter.
\vskip 4pt
\noindent
{\bf A unified universe.}
The  effective action including the matter part $S_{\textrm{m}}$ then becomes
\begin{equation}
\Gamma_{\rm eff}=
\int {\rm d}^{4}x\sqrt{-g} \left[\frac{M^2_{\rm Pl}}{2} R
- \frac{1}{\xi} R^{2}
+ m^{4}R \frac{1}{\square^2} R\right] + S_{\textrm{m}}\,,\qquad
\label{action}
\end{equation}
where $\xi$ and $m$ are phenomenological parameters which must be fixed by observations. The CMB normalization fixes $\xi \sim 1.2\times10^{-9}$ and the dark energy density today leads to $m \sim 0.3 \sqrt{H_{0} M_{\rm Pl}}$. 
This effective action is thus the minimal action (with deviations from GR of purely gravitational character) capable of unifying inflation and dark energy in a single framework {\sl sans} a cosmological constant.
In order to understand its implications for our universe, we work in a spatially flat, FRW spacetime described by the line element $ds^{2}=-dt^{2}+a^{2}(t)d{\bf x}^{2}$ where $a(t)$ is the scale factor.
Einstein's equations can  be written as $G_{\mu\nu}+\Delta G_{\mu\nu} = T_{\mu\nu}/M^2_{\rm Pl}$, where $\Delta G_{\mu\nu}$ corresponds to the correction terms in (\ref{action}). Evidently, $\Delta G_{\mu\nu}$ is covariantly conserved and its explicit form can be found in  \cite{Codello:2015pga}.
The modified Friedmann equations of motion (EOM) now read \cite{Codello:2016neo}
\begin{eqnarray}
H^{2}-\frac{12 }{M^2}\left(2H\ddot{H}+6H^{2}\dot{H}-\dot{H}^{2}\right)
-\frac{4\, m^4}{M^2_{\rm Pl}}\left(2 H^{2}S+H {\dot S}+\frac{1}{2} {\dot H} S-\frac{1}{6} {\dot U}{\dot S}\right)
&=&
\frac{\rho}{3M^2_{\rm Pl}} \label{ST_2.1}\nonumber \\
\ddot U+3H\dot U -6\left(2H^2+\dot H\right)& = & 0  \label{ST_2.2}\nonumber\\
\ddot S+3H\dot S -U& = & 0\,, \label{ST_2.3}
\end{eqnarray}
where $H={\dot a}/a$ is the Hubble parameter and the two auxiliary fields $U$ and $S$ are defined as, $U=\frac{1}{-\square}R$ and $S=\frac{1}{\square^{2}}R$, respectively.
The mass scale characterizing inflation is defined as $M^2\equiv \xi M^2_{\rm Pl}$. 
\begin{figure*}[t]
	\begin{center}
		\includegraphics[width=1.01\linewidth]{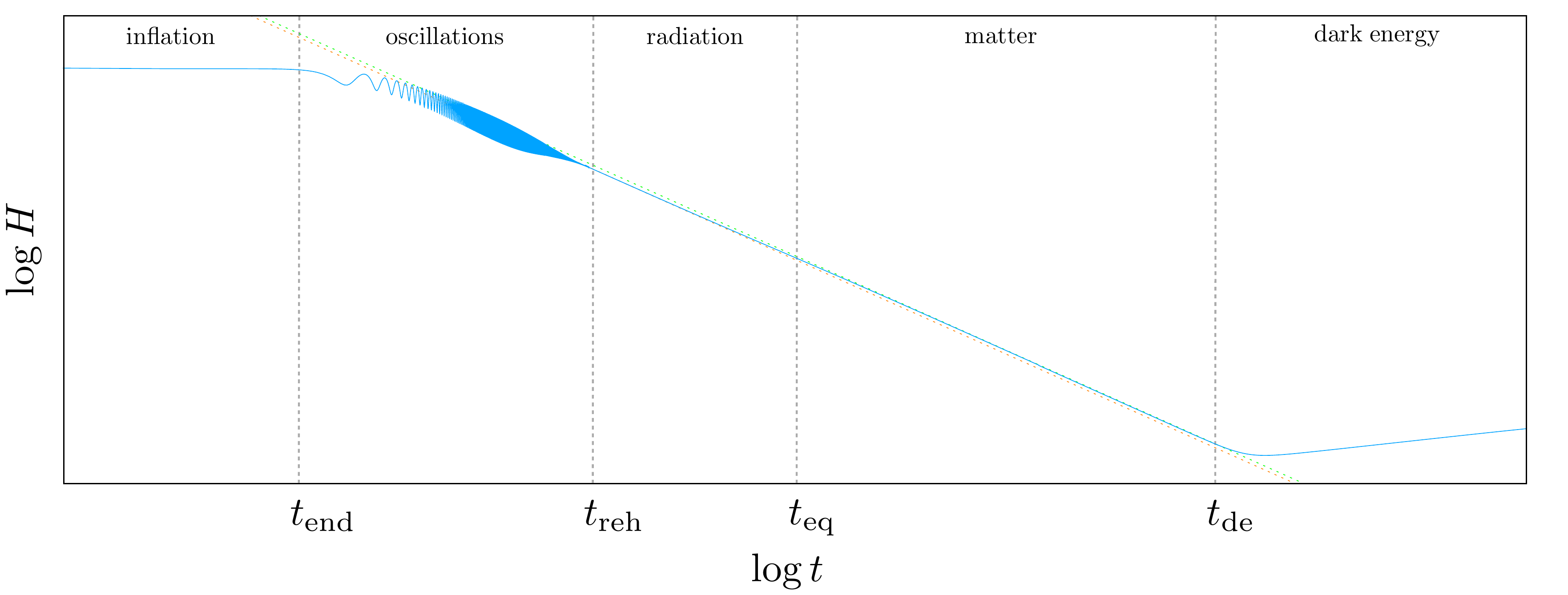}
		\vskip 2pt
		\caption{
		A unified evolution of the universe from very early times until today.
\label{universe}}
	\end{center}
\end{figure*}
The modified Friedmann EOM must be solved together with the continuity equations for radiation and matter, given by 
$\dot \rho_{\alpha}+3H(1+w_{\alpha}) \rho_{\alpha}=0,
$ where ${\alpha}=\{{\rm r},{\rm m}\}$ with $w_{\rm r}=1/3$ and $w_{\rm m}=0$.
At early times with UV corrections being most relevant, our scenario reduces to the Starobinsky model and naturally describes inflation wherein sufficient e--folds can be achieved by tuning the initial time derivate of $H$. 
The CMB normalization instead determines the value of the coupling $\xi$. 
Note that, at sufficiently early times when $R^2$ is dominant over the EH term, the theory admits a quasi de-Sitter solution leading to Starobinsky inflation. This is depicted by the constant $H$ solution on the left in Fig. \ref{universe}. Moreover, the universe in this scenario {\it gracefully} exits from inflation and enters into an oscillatory phase which is a very novel and generic feature of this scenario.  This transient oscillatory regime can be considered as the epoch of reheating and could in principle have interesting observable imprints

After the oscillatory phase is over, the universe naturally enters into the radiation dominated epoch.  After the radiation--matter equality, the universe then evolves in a matter dominated era as clearly shown in  Fig. \ref{universe}. At sufficiently late times after the equality, the non--local term $R\frac{1}{\square^2}R$, being subdominant so far, starts to become relevant and drives the current acceleration of the universe, due to the fact that $R\frac{1}{\square^2}R \to 1$ as $R/\square \to 1$ at the present epoch. 
Fig. \ref{universe}, obtained by solving the modified Friedmann EOM together with the continuity equations numerically, therefore presents a unified universe from very early times until today.

\begin{figure}[t]
\begin{center}
\includegraphics[width=0.48\linewidth]{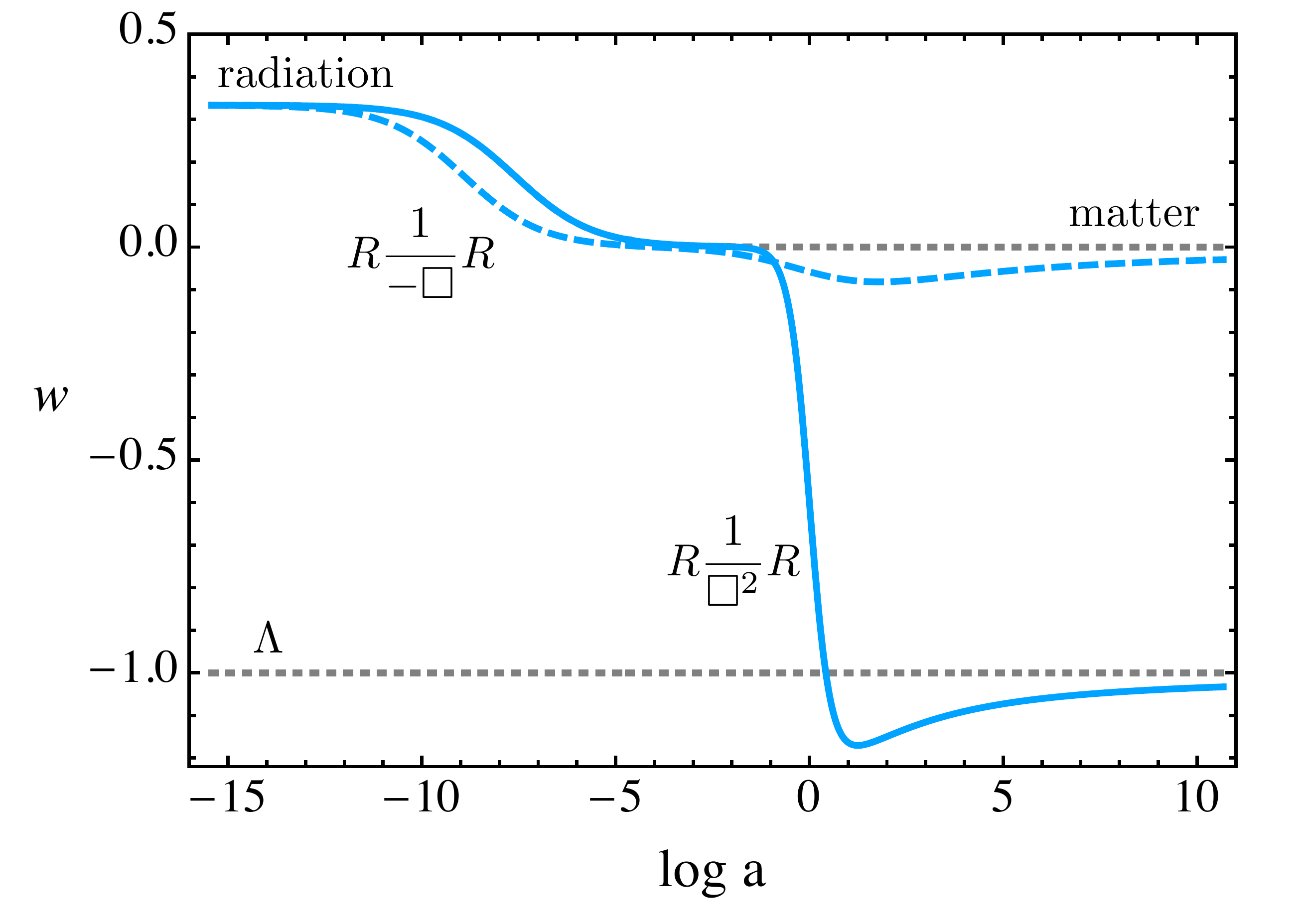}
\hskip 10pt
\includegraphics[width=0.49\linewidth]{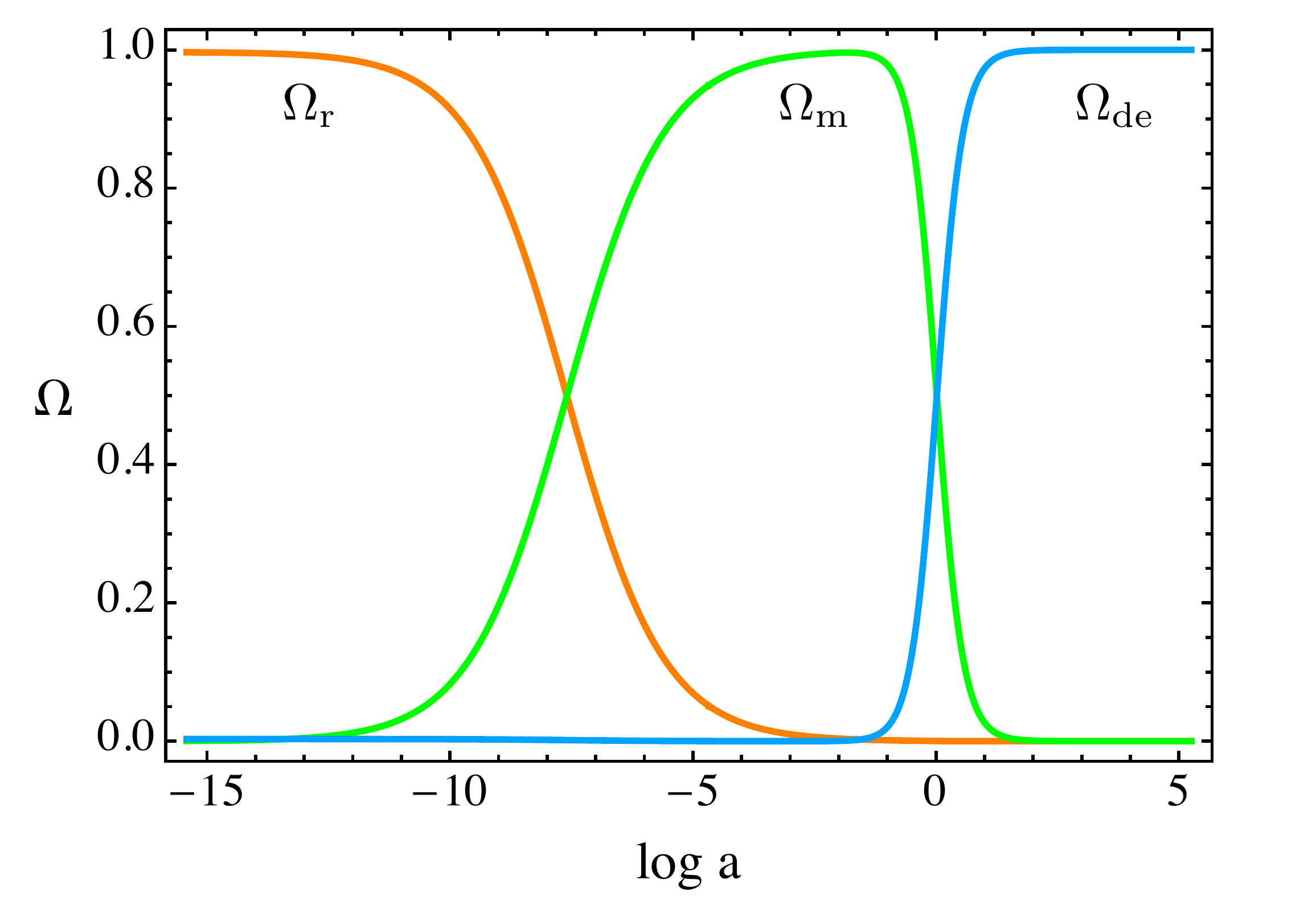}
\caption{We plot the total equation of state parameter $w$ (on left) and the dimensionless energy density ratios $\Omega_{\rm r},\Omega_{\rm m}$ and $\Omega_{\rm de}$ (on right) vs. $\log a$ from radiation epoch until today. The imprints of first two non-local terms $R\frac{1}{-\square}R$ and $R\frac{1}{\square^2}R$ are compared. It is evident that $R\frac{1}{\square^2}R$ effectively leads to $w\simeq -1$ at the present epoch. \label{w}}
\end{center}
\end{figure}

Moreover, an analytical approach is also useful in understanding the imprints of the non--local term in the background of both radiation and matter.
In terms of the conformal time $\tau=\int dt/a(t)$, the Friedmann equation including only radiation and matter can  be solved exactly as  
$a(\tau)/a_{\rm eq}=(2\sqrt{2}-2) \left(\tau/\tau_{\rm eq}\right)+(1-2\sqrt{2}+2) \left(\tau/\tau_{\rm eq}\right)^2$
with $\tau_{\rm eq} =(2\sqrt{2}-2)/a_{\rm eq} \sqrt{3 M_{\rm Pl}^2/\rho_{\rm eq}}$. 
Here, $\rho_{\rm eq}$, $a_{\rm eq}$ and $\tau_{\rm eq}$ are the energy density, scale factor and conformal time at the epoch of radiation--matter equality, respectively. 
With this solution, we have solved the Friedmann EOM corresponding to the non--local term with initial conditions $U=S=0$ and $U'=S'=0$ and treated the contributions from the non--local term when taken on the right hand side of the EOM as an effective dark energy density.
In Fig. \ref{w} on left, the total equation of state parameter $w$ vs. $\log a$ is plotted. It is evident that the universe starting from radiation epoch ($w=1/3$) smoothly transits to the matter dominated regime ($w=0$) and later evolves to the present epoch dominated by the non--local term with $w \leq -1$.
We also display the corresponding behavior for another sub--leading non--local term $R\frac{1}{-\square}R$. 
On right, we have plotted the dimensionless energy density ratios $\Omega_{\rm r}$, $\Omega_{\rm m}$ and $\Omega_{\rm de}$ vs. $\log a$. Observational consistency of $\Omega_{\rm de}\simeq0.68$ then allows us to obtain the value of the mass scale $m \sim 0.3 \sqrt{H_{0} M_{\rm Pl}}\sim 5.7 \times 10^{-4}\, {\rm eV}$.
\vskip 4pt
\noindent
{\bf Covariant EFT of gravity.}
From a phenomenological perspective, the effective action in (\ref{action}) is very appealing  as it can naturally describe all the different evolutionary phases of our universe. The immediate questions to ask are: 1) Can it have a deeper origin ? and 2) Under what assumptions is it a good low energy description of quantum gravity ?
We shall now discuss how the action in (\ref{action}) can be justified from low energy quantum gravity when it is treated by means of EFT methods.

Since we are interested in studying the evolution of the whole universe, a covariant EFT approach must be developed
in order to derive an effective action valid on an arbitrary spacetime and in particular FRW.
The leading order EFT action to the second order in curvature is given by \cite{Codello:2015mba}
\begin{equation} \label{ea_R2_Final}
\Gamma_{\rm eff} = \int\! {\rm d}^{4}x  \sqrt{-g}\left[\frac{M^2_{\rm Pl}}{2}R-\frac{1}{\xi} R^{2}
- R\, \mathcal{F}\!\left(\frac{-\square}{m^2}\right)\!R \right]\,,
\end{equation}
where the structure function $\mathcal{F}$ is completely determined once the matter content of the theory is specified. While $\xi$ is a free parameter, $m$ is in principle related to a mass scale of the underlying theory.
The structure function $\mathcal{F}$ is non--local in the low energy limit $m^2 \ll -\square$ and has the following form \cite{Codello:2015mba}
\begin{eqnarray}
\mathcal{F}\!\left(\frac{-\square}{m^2}\right)  &=&  \alpha\log\frac{-\square}{m^{2}} +\beta\frac{m^{2}}{-\square} +\gamma\frac{m^{2}}{-\square}\log\frac{-\square}{m^{2}}+\delta\!\left(\frac{m^{2}}{-\square}\right)^2+...
\label{F}
\end{eqnarray}
where the coefficients $\alpha,\beta,\gamma$ and $\delta$ are indeed {\it predictions} of the EFT of gravity which ultimately depend only on the field content of the theory \cite{Codello:2015mba}.
We recognize in (\ref{F}) the appearance of the non--local term $R\frac{1}{\square^2}R$ assumed in the model (\ref{action}).
For specific choices, one can make one or more of these coefficients vanish but in the most general case these are all non--zero and so all other non--local terms are in principle present in the effective action \cite{Codello:2015pga}.
In order to justify the model (\ref{action}) we need to show that indeed the operator $R\frac{1}{\square^2}R$ dominates the late time evolution of the universe.
For this reason in Fig. \ref{w} on left, we have compared the equation of state parameter for a dark energy fluid generated by the first two non--local terms $R\frac{1}{-\square}R$ and $R\frac{1}{\square^2}R$. As can be seen, only the latter is capable of changing the late time evolution while the first only affects the intermediate regime.
The terms containing the $\log$ operators are more delicate to define and have been discussed in \cite{Codello:2015mba}. 
The analysis shows that their behavior is sub--leading with respect to the relative operators without the $\log$ and so to first approximation can be discarded.
The $R \log \frac{-\square}{m^2} R$ term may be relevant at early times but in our model this epoch is dominated by the local $R^2$ term.
Finally, due to the smallness of the mass scale $m$, all higher order operators would be suppressed by large additional factors and to first approximation can also be discarded.
Of the terms contained in the expansion (\ref{F}) the one dominant at late times is thus the one included in the effective action (\ref{action}) and therefore, it is justifiable from first principles via EFT arguments.

Finally, we can try to answer the question: what is $m$? For scalars and fermions, $m$ is the effective mass of the particle and is required to be $m \sim 0.57\, {\rm meV}$ for observational consistency, making it a very light particle. Photons do not  contribute while gravitons can induce an effective gravitational mass $m^2_{\rm grav}=2\, V(v)/M^2_{\rm Pl}$ (where $V(\varphi)$ is the scalar potential with $v$ its minimum). Such a light particle must have been relativistic all its life during the evolution of the universe and would contribute to the effective number of relativistic degrees of freedom. 
The presence of such extra degrees of freedom can be tightly constrained using Planck and other observations \cite{Ade:2015xua}. 
Structure formation can also be used to constrain the imprints of such light particles. 
\vskip 4pt
\noindent
{\bf Discussion.}
The simple yet elegant scenario presented here exhibits a unified evolution of the universe, starting from an inflationary regime at very early times to the dark energy phase at the present epoch. The framework is not only phenomenological but can also be justified using an effective approach to low energy quantum gravity. The mass scale $m$ can either be identified with the mass of a very light species or with the effective gravitational mass. This connection is not only satisfactory from a theoretical point of view but it also provides a way to reduce the number of free parameters by linking $m$ to the underlying theory.  To conclude, this scenario represents a unified cosmological model with one parameter less, and is thus characterized by a higher predictive power that can in principle be falsified. 
\vskip 4pt
\noindent
{\bf Acknowledgements.}
The CP$^3$-Origins centre is partially funded by the Danish National Research Foundation, grant number DNRF90.

%%%%%%%%%%%%%%%%%%%%%%%%%%%%%%%%%%%%%%%%%%%%%%%%%%%%%%%%%%%%%%%%%%%%%%%%%%%%%%%
\bibliographystyle{JHEP}
\bibliography{EFT_GR_short_Bibliography}

%%%%%%%%%%%%%%%%%%%%%%%%%%%%%%%%%%%%%%%%%%%%%%%%%%%%%%%%%%%%%%%%%%%%%%%%%%%%%%%

\end{document}